# Robust dynamic mitigation of instabilities

S. Kawata, and T. Karino

Graduate School of Engineering, Utsunomiya University, 321-8585 Utsunomiya, Japan

A dynamic mitigation mechanism for instability growth was proposed and discussed in the paper [Phys. Plasmas **19**, 024503 (2012)]. In the present paper the robustness of the dynamic instability mitigation mechanism is discussed further. The results presented here show that the mechanism of the dynamic instability mitigation is rather robust against changes in the phase, the amplitude and the wavelength of the wobbling perturbation applied. Generally instability would emerge from the perturbation of the physical quantity. Normally the perturbation phase is unknown so that the instability growth rate is discussed. However, if the perturbation phase is known, the instability growth can be controlled by a superposition of perturbations imposed actively: if the perturbation is induced by, for example, a driving beam axis oscillation or wobbling, the perturbation phase could be controlled and the instability growth is mitigated by the superposition of the growing perturbations.

PACS number: 52.57.Fg, 41.85.-p, 52.58.Hm

I. INTRODUCTION

In general it would be hard to control instabilities in plasma and fluid. Usually the instability growth rate is employed to examine the plasma state, and the stabilization mechanisms are also studied. For example, the dynamic stabilization for the Rayleigh-Taylor instability (RTI) [1-6] has been studied for a uniform compression [7, 8] of a fusion fuel target in inertial confinement fusion. The RTI dynamic stabilization was proposed many years ago [1, 2] in inertial fusion; the oscillation amplitude of the driving acceleration should be sufficiently large to stabilize RTI [1-6]. In inertial fusion, the fusion fuel compression is essentially important to reduce an input driver energy [7, 8], and the implosion uniformity is one of critical issues to release the fusion energy stably [9, 10].

Instability grows from a perturbation of the physical quantity, and the perturbation phase is unknown in plasmas. Therefore, usually the instability growth rate is focused and examined. On the other hand, in an unstable system there is a well-known feedback control mechanism in which the perturbation amplitude and phase are detected, and the growing perturbation is compensated by the active feedback control. However, in plasmas and fluids it is difficult to measure the instability phase and amplitude, and so the perfect active feedback control cannot be realized.

If we actively impose the perturbation phase by the driving energy source wobbling or oscillation, and so if we know or define the phase of the perturbations imposed actively, the perturbation growth can be controlled in a similar way [11-13] as the feedback control mechanism. For example, the two-stream instability growth would be controlled by a time-dependent drift velocity of the driving beam [14]. When the driving beam longitude velocity is oscillating, the two-stream instability perturbation phase changes in time. At each time the driving beam introduces a new perturbation phase, and the actual instability growth is defined by the superposition of all the growing perturbations by the time-dependent drift velocity. Another example is found in the filamentation instability: the growth of the filamentation instability [15-19], for example, driven by a particle beam could be controlled by the beam axis oscillation or wobbling. The oscillating beam induces the perturbations at each time and also could define the perturbation phase. Therefore, the successive phase-defined perturbations are superposed, and the actual instability growth would be controlled or mitigated. Another interesting example can be found in heavy ion beam inertial fusion; the heavy ion accelerator could provide a beam axis wobbling with a high frequency[20-22]. The wobbling heavy ion beams also define the perturbation phase. This means that the perturbation phase is known, and so the perturbations successively imposed are superposed in the plasma. The heavy ion beams accelerate the fusion target fuel with a large acceleration in inertial fusion. The wobbling heavy ion beams would provide a small oscillating acceleration perturbation in an inertial fusion fuel target during the target implosion. So the RTI growth would be reduced by the phase-controlled superposition of perturbations in heavy ion inertial fusion[11-13].

In this paper we discuss the robustness of the dynamic mitigation mechanism for instabilities presented in Refs. 11-13. The results presented here show that the mechanism of the dynamic instability mitigation is rather robust against changes in the phase, amplitude and wavelength of the wobbling perturbation applied. The promising results presented in this paper ensure the viability of the mechanism of the dynamic instability mitigation.

## II. DYNAMIC INSTABILITY MITIGATION

Let us consider an unstable system, which has one mode of $a = a_0 e^{ikx+\gamma t}$. Here $a_0$ is the amplitude, $k = 2\pi/\lambda$ is the wave number, $\lambda$ the wave length and $\gamma$ the growth rate of the instability. An example initial perturbation is shown in Fig. 1(a). At $t$=0 the perturbation is imposed. The initial perturbation grows with $\gamma$. After $\Delta t$, if another perturbation, which has an inverse phase, is actively imposed (see Fig. 1(b)), the overall amplitude is the superposition of all the perturbations, and so the actual perturbation amplitude is well mitigated as shown in Fig. 1(c). This is an ideal example for the dynamic instability mitigation[11-13].

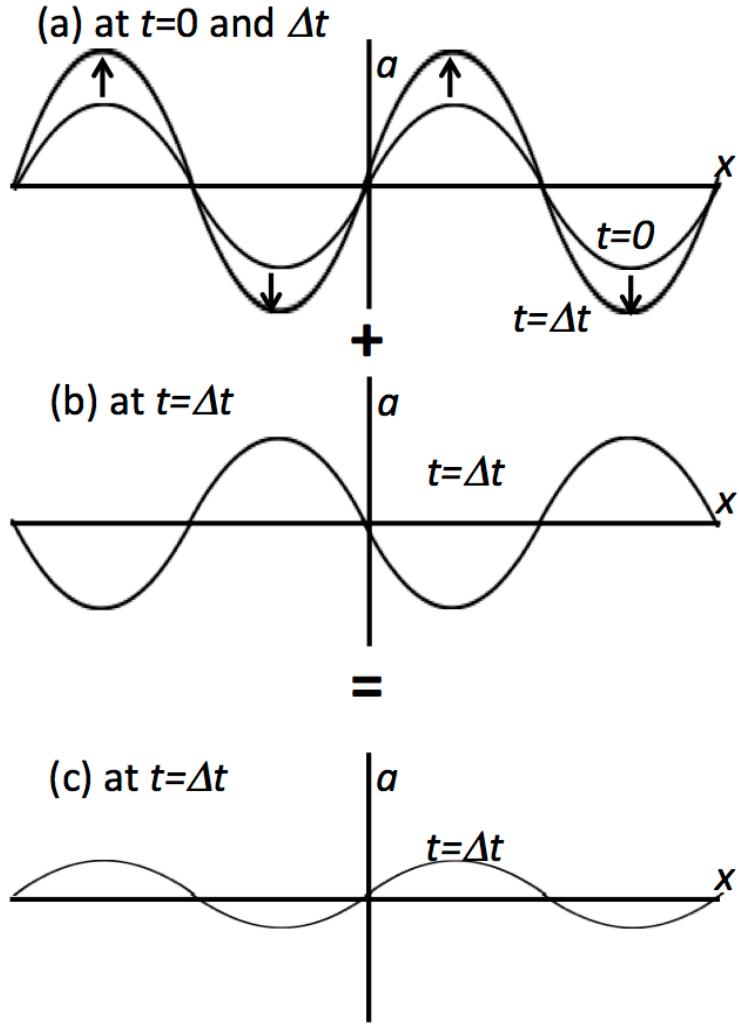

FIG. 1 An ideal example concept of the dynamic mitigation. (a)At $t$=0 a perturbation is imposed. The initial perturbation grows with $\gamma$. (b) After $\Delta t$ another perturbation, which has an inverse phase, is actively imposed, so that (c) the actual perturbation amplitude is mitigated very well after the superposition of the initial and additional perturbations.

In plasmas the perturbation phase and amplitude cannot be measured. So the perfect feedback control cannot be realized in plasmas and fluids. However, an electron beam can provide its axis wobbling motion or a time-dependent modulation of the beam velocity. A heavy ion beam accelerator can also provide a controlled wobbling or oscillating beam with a high frequency[20, 21]. They would provide the defined phase and amplitude of perturbations.

When the instability driver wobbles uniformly in time, the imposed perturbation for a physical quantity of $F$ at $t=\tau$ may be written as

$$F = \delta F e^{i\Omega\tau} e^{\gamma(t-\tau)+i\vec{k}\cdot\vec{x}}. \quad (1)$$

Here $\delta F$ is the amplitude, $\Omega$ the wobbling or oscillation frequency defined actively by the driving wobbler, and $\Omega\tau$ the phase shift of superimposed perturbations. At each time $t$, the wobbler or the modulated driver provides a new perturbation with the phase and the amplitude actively defined by the driving wobbler itself. The superposition of the perturbations provides the actual perturbation at $t$ as follows:

$$\int_0^t d\tau \; \delta F e^{i\Omega\tau} e^{\gamma(t-\tau)+i\vec{k}\cdot\vec{x}} \propto \frac{\gamma+i\Omega}{\gamma^2+\Omega^2} \delta F e^{\gamma t} e^{i\vec{k}\cdot\vec{x}} \quad (2)$$

When $\Omega \gg \gamma$, the perturbation amplitude is reduced by the factor of $\gamma/\Omega$, compared with the pure instability growth ($\Omega=0$) based on the energy deposition nonuniformity[11, 12]. When $\Omega \cong \gamma$, the amplitude mitigation factor is still about 50%. The result in Eq. (2) presents that the perturbation phase should oscillate with $\Omega \gtrsim \gamma$ for the effective amplitude reduction.

Figure 2 shows an example simulation for RTI, which has a single mode. In this example, two stratified fluids are superimposed under an acceleration of $g = g_0 + \delta g$. The density jump ratio between the two fluids is 10/3. In this specific case the wobbling frequency $\Omega$ is $\gamma$, the amplitude of $\delta g$ is $0.1g_0$, and the results shown in Figs. 2 are those at $t = 5/\gamma$. In Fig. 2(a) $\delta g$ is constant and drives the RTI as usual, and in Fig. 2(b) the phase of $\delta g$ is shifted or oscillates with the frequency of $\Omega$ as stated above for the dynamic instability mitigation. The RTI growth mitigation ratio is 72.9% in Fig. 2. The growth mitigation ratio is defined by ($H_0$ - $H_{mitigate}$)/$H_0$×100%. Here *H* is defined as shown in Fig. 2(a), $H_0$ shows the deviation amplitude of the two-fluid interface in the case in Fig. 2(a) without the oscillation ($\Omega = 0$), and $H_{mitigate}$ presents the deviation for the other cases with the oscillation ($\Omega \neq 0$). The example simulation results support well the effect of the dynamic mitigation mechanism.

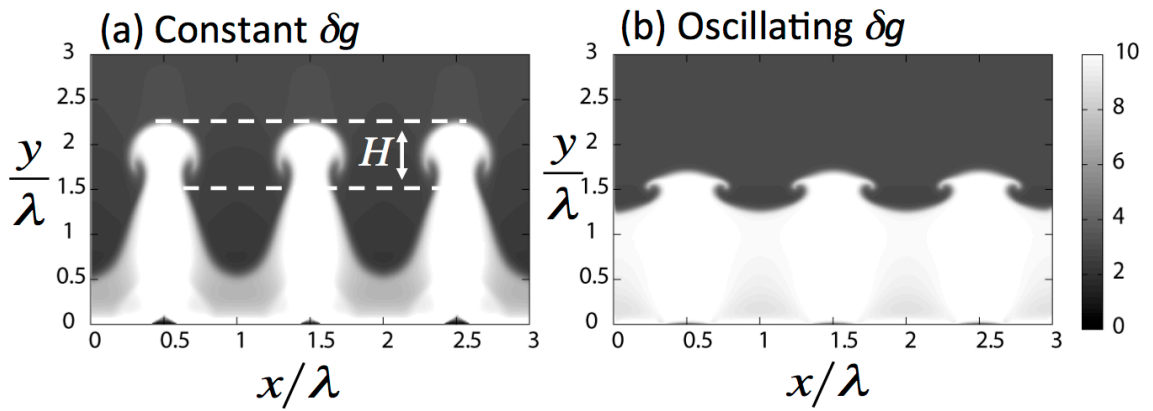

FIG. 2 Example simulation results for the Rayleigh-Taylor instability (RTI) mitigation. $\delta g$ is 10% of the acceleration $g_0$ and oscillates with the frequency of $\Omega=\gamma$. As shown above and in Eq. (2), the dynamic instability mitigation mechanism works well to mitigate the instability growth.

III. ROBUSTNESS OF DYNAMIC MITIGATION OF INSTABILITY

In order to check the robustness of the dynamic instability mitigation mechanism, here we study the effects of the change in the phase, the amplitude and the wavelength of the wobbling perturbation $\delta F$, that is, $\delta g$ in Fig. 2 on the dynamic instability mitigation.

When the perturbation amplitude $\delta F = \delta F(t)$ depends on time or oscillates slightly in time, the dynamic mitigation mechanism is examined first. We consider $\delta F(t) = \delta F_0 \left(1 + \Delta e^{i\Omega' t}\right)$ in Eq. (1). Here $\Delta \ll 1$. In this case, Eq. (2) is modified as follows:

$$\int_0^t d\tau \; \delta F e^{i\Omega\tau} e^{\gamma(t-\tau)+i\vec{k}\cdot\vec{x}} \propto \left\{\frac{\gamma+i\Omega}{\gamma^2+\Omega^2} + \Delta \frac{\gamma+i(\Omega+\Omega')}{\gamma^2+(\Omega+\Omega')^2}\right\} \delta F_0 e^{\gamma t} e^{i\vec{k}\cdot\vec{x}} \quad (3)$$

When $\Delta \ll 1$ in Eq. (3), just a minor effect appears on the dynamic mitigation of the instability.

We also performed the fluid simulations. In the simulations $\delta F = \delta g(1 - \Delta sin\Omega' t)$. The RTI is simulated again based on the same parameter values shown in Fig. 2 except the perturbation amplitude oscillation $\delta F(t)$. In the simulations we employ $\Omega'$ =3 $\Omega$, $\Omega$ and $\Omega$/3 in Eq. (3). For $\Delta$=0.1 and 0.3, and for $\Omega'$ =3 $\Omega$, $\Omega$ and $\Omega$/3, the RTI growth reduction ratio is 54.9~73.2% at $t = 5/\gamma$. The results by the fluid simulations and Eq. (3)

demonstrate that the perturbation amplitude oscillation $\delta F = \delta F(t)$ is uninfluential as long as $\Delta \ll 1$.

When the oscillation frequency $\Omega$ of the perturbation $\delta F$ depends on time ($\Omega = \Omega(t)$), the time-dependent frequency means that $\Omega(t)$ would consist of multiple frequencies: $e^{i\Omega t} = \sum_i \Delta_i e^{i\Omega_i t}$. In this case Eq. (3) becomes

$$\int_0^t d\tau \ \delta F e^{i\Omega \tau} e^{\gamma(t-\tau)+i\vec{k}\cdot\vec{x}} \propto \sum_i \Delta_i \frac{\gamma+i\Omega_i}{\gamma^2+\Omega_i^2} \delta F e^{\gamma t} e^{i\vec{k}\cdot\vec{x}}. \tag{4}$$

The result in Eq. (4) shows that the highest frequency of $\Omega_i$ contributes to the instability mitigation. In a real system the highest frequency would be the original wobbling frequency $\Omega$ or so, and the largest amplitude of $\Delta_i$ is also that for the original wobbling mode. So when the frequency change is slow, the original wobbler frequency of $\Omega$ contributes to the mitigation.

The fluid simulations are also done for the RTI with $\Omega(t) = \Omega\left(1 + \Delta \sin\Omega' t\right)$ together with the same parameter values employed in Fig. 2. In this case $\Delta$=0.1 and 0.3, and $\Omega'$ =3 $\Omega$, $\Omega$ and $\Omega$/3. The growth reduction ratio was 66.9~74.0% at $t = 5/\gamma$. The little oscillation of the imposed perturbation oscillation frequency $\Omega(t)$ has a minor effect on the dynamic instability mitigation.

When the wobbling wavelength $\lambda = 2\pi/k$ depends on time, one can expect as follows in a real system: $k(t) = k_0 + \Delta k e^{i\Omega'_k t}$ and $k_0 \gg \Delta k$. In this case the wobbling wavelength changes slightly in time, and Eq. (3) becomes as follows:

$$\int_0^t d\tau \ \delta F e^{i\Omega\tau} e^{\gamma(t-\tau)+ik\cdot x} \propto \delta F e^{\gamma t + ik_0 \cdot x} \int_0^t d\tau \ e^{(i\Omega-\gamma)\tau} \sum_{m=-\infty}^{\infty} i^m J_m(\Delta k \cdot x) e^{im\Omega'_k \tau}$$

$$\propto \sum_{m=-\infty}^{\infty} i^m J_m(\Delta k \cdot x) \int_0^t d\tau \ e^{i(\Omega+m\Omega'_k)\tau - \gamma\tau} \propto \sum_{m=-\infty}^{\infty} i^m J_m(\Delta k \cdot x) \frac{\gamma + i(\Omega+m\Omega'_k)}{\gamma^2+(\Omega+m\Omega'_k)^2} \tag{5}$$

Here $J_m$ is the Bessel function of the first kind. The result in Eq. (5) demonstrates that the instability growth reduction effect is not degraded by the small change in the wobbling wavelength. In actual situations the mode $m = 0$ contributes mostly to the instability mitigation, and in this case the original reduction effect shown in Eq. (2) is recovered.

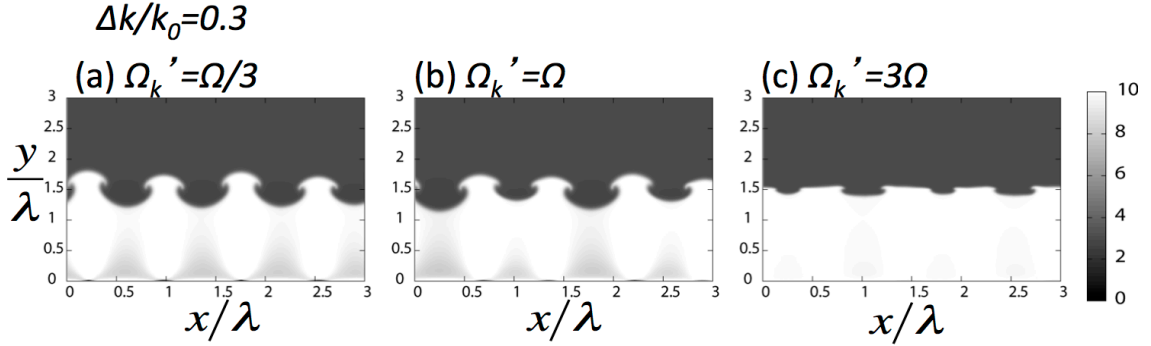

FIG. 3 Fluid simulation results for the RTI mitigation for the time-dependent wobbling wavelength $k(t) = k_0 + \Delta k e^{i\Omega'_k t}$ at $t = 5/\gamma$. In the simulations $\Delta k/k_0 = 0.3$, and (a) $\Omega'_k = \Omega/3$, (b) $\Omega'_k = \Omega$ and (c) $\Omega'_k = 3\Omega$. The dynamic mitigation mechanism is also robust against the time change of the perturbation wavelength $k(t)$.

The fluid simulations are also performed for this case $k(t) = k_0 + \Delta k e^{i\Omega'_k t}$. Figure 3 shows the example simulation results for $\Delta k/k_0 = 0.3$ and $\Omega'_k = 3\Omega$, $\Omega$ and $\Omega/3$. Figure 3(a) shows the RTI growth reduction ratio of 61.3% for $\Omega'_k = \Omega/3$, Fig. 3(b) shows 68.0% for $\Omega'_k = \Omega$ and Fig. 3(c) shows 93.3% for $\Omega'_k = 3\Omega$ at $t = 5/\gamma$. For a realistic situation $\Omega'_k \sim \Omega$, where $\Omega$ is the wobbling or modulation frequency.

All the results shown above demonstrate that the dynamic instability mitigation mechanism proposed is rather robust against the changes in the amplitude, the phase and the wavelength of the wobbling or modulating perturbation of $\delta F$ in general or $\delta g$ in RTI.

IV. SUMMARY

We have discussed the dynamic mitigation method, in which the perturbation growth is controlled and mitigated by the wobbler or the driver modulation. In this paper we focus on the robustness of the dynamic mitigation mechanism of the instability. The changes in the perturbation frequency, amplitude and wavelength were examined. The theoretical and simulation results demonstrate that the dynamic instability mitigation mechanism is rather robust against the changes in the perturbation frequency, the amplitude and the wavelength of the driver wobbling motion or the driver modulation. The results in this paper show the viability of the mechanism of the dynamic instability mitigation.

The wobbling or the modulation defines the imposed perturbation phase and amplitude at each time. Consequently the overall superposition of all the perturbations mitigates its growth through the control of the wobbling or modulating motion. The dynamic mitigation would work for the mitigation of instabilities in which the imposed perturbation phase is controlled actively.

For RTI, the growth rate $\gamma$ is larger for a short wavelength. If $\Omega \ll \gamma$, the modes cannot be mitigated. In addition, if there are other sources of perturbations in the physical system and if the perturbation phase and amplitude cannot be controlled at all, the dynamic mitigation mechanism proposed here does not work. For example, when the shell thickness of an inertial fusion fuel target is not uniform at the target fabrication process, the dynamic mitigation mechanism does not work. In this sense the dynamic mitigation mechanism is not almighty. Especially for a uniform compression of an inertial fusion fuel all the instability stabilization and mitigation mechanisms would contribute to release the fusion energy. How to produce the controlled phase shift should be also studied further. As mentioned above, a wobbling heavy ion beam or an oscillating electron beam or a time-dependent driver velocity modulation provides a good example phase controller. A short-pulse diver bunch train, for example, a short-pulse-laser train or a particle beam train would also provide a phase controller. The phase-control method should be studied further.

ACKNOWLEDGEMENTS
This work is partly supported by MEXT, JSPS, ILE/Osaka Universiy,